\documentclass{sty/IEEEtran}
\usepackage{graphicx}
\usepackage{amsmath}
\usepackage{amssymb}
\usepackage{enumitem}
\usepackage{booktabs}
\usepackage{url}
\usepackage{algorithm}
\usepackage{algpseudocode}

\begin{document}
\title{Time-Domain Data-Driven Loewner Reduced-Order Models for Electromagnetic Simulation}
\author{Pablo Biurrun, Carlos J. Ramos-Salas, Luis D. Angulo and Valent\'{i}n de la Rubia
\thanks{This work was supported in part by the European Innovation Council under the EIC Pathfinder Open 2022 program through the project \emph{Plasma reconfigUrabLe metaSurface tEchnologies} (PULSE) (protocol number 101099313), in part by the European Union through the project iSENSE -- In-Situ Monitoring of Electromagnetic Interference (Grant Agreement No.~101168880), in part by the European Union under Grant Agreement No.~101101961 (HECATE), and in part by the project ESAMA (PID2022-137495OB-C31), funded by MICIU/AEI/10.13039/501100011033.}
\thanks{Pablo Biurrun and Valent\'{i}n de la Rubia are with the Departamento de Matem\'{a}tica Aplicada a las {TIC}, ETSI de Telecomunicaci\'{o}n, Universidad Polit\'{e}cnica de Madrid, 28040 Madrid, Spain (e-mails: p.biurrun@upm.es; valentin.delarubia@upm.es). Carlos J. Ramos-Salas and Luis D. Angulo are with the Group of Electromagnetism of Granada (GEG), in University of Granada, Spain (e-mails: cramoss@ugr.es; lmdiazangulo@ugr.es)}
}
\maketitle
\begin{abstract}
This paper presents a data-driven Loewner framework for constructing reduced-order models (ROMs) from time-domain electromagnetic (EM) simulation data. High-fidelity simulation of microwave devices can be computationally expensive, particularly when fine spatial discretization and long observation windows are required. The proposed method uses transient input–output data to construct Loewner and shifted Loewner matrices, from which a reduced descriptor realization is obtained through singular value decomposition (SVD). The resulting models are identified from a limited training window and used to predict responses over substantially longer time intervals, with prediction windows up to 50 times the duration of the training window. The method is formalized as an explicit, reproducible algorithm and validated against full-order simulations for four microwave structures: a substrate-integrated-waveguide filter, a frequency-selective surface, an evanescent-mode coaxial comb-line filter, and a diplexer. The results demonstrate the potential of Loewner-based identification to provide compact models of electromagnetic responses while reducing the need for repeated full-order simulation.
\end{abstract}


\IEEEpeerreviewmaketitle

\section{Introduction}

\PARstart{M}{icrowave} and millimeter-wave devices, including compact substrate-integrated-waveguide filters, three-dimensional frequency-selective surfaces, evanescent-mode coaxial filters, closely spaced channel diplexers, and direct-coupled waveguide cavity filters \cite{moscato2016,parellada2024,bastioli2024,lamecki2014,bastioliTMTT2024}, rely on electromagnetic interactions across multiple spatial scales. Their performance, including resonance locations, passband characteristics, and channel isolation, can be difficult, or even impossible, to predict accurately using circuit intuition or closed-form analysis alone. Full-wave electromagnetic simulation is therefore often necessary to capture the behavior of the complete structure.

Full-wave numerical simulation is consequently not a convenience but the primary design instrument. The finite-difference time-domain method \cite{taflove2005} and finite-element time-domain schemes \cite{jin2014} solve Maxwell's equations directly on a discretization of the structure and, given a fine enough mesh, reproduce these coupled phenomena faithfully. However, there is a caveat which is the great cost at which high-fidelity results come. Resolving multi-scale geometry drives the number of spatial degrees of freedom into the hundreds of thousands or millions. On the other hand, accurately resolving a high-$Q$ response, whose transient rings for a long time relative to a period, drives the number of time steps into the tens or hundreds of thousands. The total cost is the product of these two, so a single narrowband filter or diplexer can occupy a workstation for hours while a platform-level electromagnetic-compatibility problem can occupy a cluster for days. In a design loop that must be traversed dozens of times, this is prohibitively costly.
 
A classical response to this difficulty is model order reduction (MOR), which replaces the large discretized system by a much smaller one that reproduces the input--output behavior of interest \cite{antoulas2005,benner2015,mrozowski2001}. The projection-based branch of this field is mature. Balanced truncation identifies dominant states through their joint controllability and observability \cite{moore1981}; Krylov-subspace and moment-matching methods build a reduced model by matching transfer-function moments at selected expansion points, an idea developed for circuit simulation through Pad\'e-via-Lanczos \cite{feldmann1995} and made provably passive for RLC systems by PRIMA \cite{odabasioglu1998}, with the numerical-linear-algebra foundations surveyed in \cite{bai2002}; and reduced-basis methods construct a problem-specific approximation space and can provide certified error bounds under suitable assumptions, an approach that has been applied successfully to fast frequency sweeps in microwave devices \cite{delarubia2009,delarubia2018} and, more recently, to the time-domain finite-element setting itself \cite{medeiros2024,medeiros2025}. These methods are powerful, but conventional projection-based approaches generally require access to the discretized system matrices or the ability to perform operations on the underlying operators. For commercial solvers  that provide only input--output responses, or for problems where such operations are computationally prohibitive, this requirement can be difficult to satisfy.
 
This is the point at which state-space system identification (SSSI) becomes relevant as a framework. Rather than projecting a known operator onto a smaller subspace, SSSI asks a different question: given only excitation and response data, can a state-space realization $(E,A,B,C)$ \cite{antoulas2005} be constructed that reproduces the observed responses and predicts the system's behavior under unseen conditions? The solver is treated as a black box, requiring only input–output data rather than access to its internal discretized operators. The idea originates in systems and control theory and has accumulated a rich set of algorithms, several of which are worth distinguishing because they make different compromises.
 
Realization-based methods start from Hankel matrices constructed from the impulse response: the eigensystem realization algorithm (ERA) uses their singular value decomposition (SVD) to construct a reduced-order realization, with the singular-value spectrum providing guidance for model-order selection \cite{juang1985}. Subspace identification extends this approach to noisy input--output records by estimating the system's state subspace through projections of block-Hankel data matrices, as in N4SID \cite{vanoverschee1994} and MOESP-type methods \cite{verhaegen1992}. A second family operates in the frequency domain and constructs rational approximations through iterative least-squares procedures, beginning with the Sanathanan--Koerner (SK) iteration \cite{sanathanan1963} and extending to vector fitting (VF) \cite{gustavsen1999,mrozowski2008}. The convergence behavior and $\mathcal{H}_2$ interpretation of vector fitting have since been studied in greater detail \cite{drmac2015}, while its practical application in passive macromodeling is documented in \cite{grivettalocia2016}. A third family uses adaptive barycentric rational approximation, as in the AAA algorithm \cite{nakatsukasa2018}, which greedily selects support points based on the residual error at the sampled frequencies. These approaches differ in how they select model order, represent system dynamics, and manage numerical stability. In particular, ERA and subspace identification generally require an order estimate, whereas vector fitting typically requires a prescribed number of poles and AAA adaptively increases the approximation degree according to a stopping criterion. Although these methods offer efficient alternatives to direct full-order simulation, their accuracy and reliability depend on data quality, model selection, and the numerical behavior of the identification algorithm.
 
The Loewner framework occupies a distinctive position among these alternatives. Introduced by Mayo and Antoulas to address the generalized realization problem \cite{mayo2007}, it constructs a descriptor realization directly from sampled transfer-function data using two structured matrices: the Loewner matrix and its shifted counterpart, whose entries are formed from divided differences of the data. Unlike optimization-based fitting methods, the basic construction is algebraic and does not require a nonlinear optimization procedure. The initial realization order is determined by the available data, while singular-value analysis of the Loewner matrices or associated pencil can guide the selection of a lower model order. The framework was subsequently adapted to multiport macromodeling from frequency-domain measurements \cite{lefteriu2010}, extended to parametrized systems \cite{ionita2014}, and developed further within the broader theory of interpolatory model reduction \cite{antoulas2020}. Particularly relevant to the present work, Peherstorfer, Gugercin, and Willcox showed that this framework can also be applied non-intrusively to time-domain input--output trajectories by inferring frequency-response data from transient simulations and using these data to construct a Loewner realization \cite{peherstorfer2017}. This provides a basis for applying Loewner-based model reduction to data generated by time-domain electromagnetic solvers, without requiring access to their internal discretized operators.

This paper develops this approach into a practical workflow for electromagnetic simulation. It presents the Loewner and shifted Loewner matrices explicitly as an $n\times n$ pair and formalizes the complete identification and prediction procedure in a single implementation-level algorithm. The procedure encompasses transient-data processing, spectral estimation and masking, construction of a reduced descriptor realization, and prediction beyond the training interval. The method is validated against full-order reference simulations from in-house finite-element solvers on the four electromagnetic structures introduced above. These cases span full-order model (FOM) sizes from about 136.000 to 479.000 degrees of freedom, with excitation bandwidths ranging from a few hundred megahertz to 24 GHz. The influence of the training-window duration is investigated explicitly, and the resulting models are shown to reproduce the full-order responses over prediction windows up to fifty times the training-window duration. Finally, the paper outlines possible extensions to nonlinear dynamical systems through Koopman-operator approaches, including dynamic mode decomposition (DMD) \cite{schmid2010,rowley2009}.

\section{Loewner Method}
\label{sec:method}

\subsection{Frequency-Domain Data and Loewner Matrices}

Consider a linear time-invariant (LTI) system for which time-domain
input and output snapshots, $u[k]$ and $y[k]$, are available at a
sampling interval $\delta t$, where $k$ represents each particular time-step. The fast Fourier transform (FFT) is
applied to the signals over a training window of duration
$t_{\mathrm{train}}$, yielding the frequency-domain representations
$U_{\mathrm{train}}(j\omega_k)$ and
$Y_{\mathrm{train}}(j\omega_k)$. At frequency bins where the input
spectrum has sufficient magnitude, the transfer-function samples
are estimated as
\begin{equation}
    H(j\omega_k)
    =
    \frac{Y_{\mathrm{train}}(j\omega_k)}
         {U_{\mathrm{train}}(j\omega_k)}.
    \label{eq:frequency_response_estimate}
\end{equation}
Frequencies are restricted to the band sufficiently excited by the input. The retained frequency-response samples are partitioned into two
disjoint sets of equal size $n$. The right interpolation data are
denoted by $(\mu_i,H_{\mu_i})$, $i=1,\ldots,n$, and the left
interpolation data by $(\lambda_j,H_{\lambda_j})$,
$j=1,\ldots,n$. Here, $\mu_i$ and $\lambda_j$ are distinct
interpolation points in the complex $s$-plane. The Loewner matrix
$\mathbb{L}$ and the shifted Loewner matrix $\mathbb{L}_s$ are defined as
\begin{equation}
    \mathbb{L}_{ij}
    =
    \frac{H_{\mu_i}-H_{\lambda_j}}
         {\mu_i-\lambda_j},
    \label{eq:loewner_matrix}
\end{equation}
and
\begin{equation}
    (\mathbb{L}_s)_{ij}
    =
    \frac{\mu_i H_{\mu_i}-\lambda_j H_{\lambda_j}}
         {\mu_i-\lambda_j},
    \label{eq:shifted_loewner_matrix}
\end{equation}
for $i,j=1,\ldots,n$. Both matrices belong to
$\mathbb{C}^{n\times n}$.

\subsection{Reduced-Order Realization}

The Loewner framework constructs a reduced-order realization
directly from frequency-response data, without requiring explicit
access to the full-order system matrices. The resulting model is
represented in descriptor form as
\begin{equation}
    E_r \dot{x}(t)
    =
    A_r x(t)+B_r u(t),
    \qquad
    y(t)=C_r x(t),
    \label{eq:descriptor_model}
\end{equation}
where $x(t)\in\mathbb{C}^{r}$ is the reduced state and $r$ is the
reduced model order. For a single-input, single-output system,
$E_r,A_r\in\mathbb{C}^{r\times r}$,
$B_r\in\mathbb{C}^{r\times 1}$, and
$C_r\in\mathbb{C}^{1\times r}$. The descriptor representation
does not require $E_r$ to be invertible, provided that the
resulting system is well posed.

Using the sign convention adopted here, the reduced transfer
function is
\begin{equation}
    H_r(s)
    =
    C_r(sE_r-A_r)^{-1}B_r.
    \label{eq:reduced_transfer_function}
\end{equation}

No separate feedthrough term is required: in the Loewner realization any direct-feedthrough component of $H(s)$ is absorbed into the descriptor pencil \cite{mayo2007}, so $D=0$ throughout.

The reduced-order realization is obtained by projecting the Loewner pencil
onto its dominant left and right subspaces \cite{mayo2007}, \cite{antoulas2020}. These are computed from
the singular value decompositions of the horizontally and vertically stacked
Loewner matrices,
\begin{equation}
\begin{bmatrix} \mathbb{L} & \mathbb{L}_s \end{bmatrix}
  = Y \Sigma_{\ell} \tilde{X}^{H}, \qquad
\begin{bmatrix} \mathbb{L} \\ \mathbb{L}_s \end{bmatrix}
  = \tilde{Y} \Sigma_{r} X^{H},
\label{eq:svd}
\end{equation}
where $Y, X \in \mathbb{C}^{n \times n}$ contain the left and right singular
vectors used for projection. For a reduced order $r$, the first $r$ columns of
$Y$ and $X$ are retained, defining $Y_r, X_r \in \mathbb{C}^{n \times r}$. The
reduced realization matrices are then
\begin{align}
E_r &= -Y_r^{H} \mathbb{L} X_r,   \label{eq:Er}\\
A_r &= -Y_r^{H} \mathbb{L}_s X_r, \label{eq:Ar}\\
B_r &= Y_r^{H} W,                 \label{eq:Br}\\
C_r &= V^{T} X_r,                 \label{eq:Cr}
\end{align}
where $W = [H_{\mu_1}, \dots, H_{\mu_n}]^{T}$ and
$V = [H_{\lambda_1}, \dots, H_{\lambda_n}]^{T}$ denote the right and left
interpolation-value vectors, respectively. The order $r$ is selected as the
smallest integer for which the cumulative energy of the singular values
$\sigma_i$ of $[\,\mathbb{L}\;\; \mathbb{L}_s\,]$ exceeds a prescribed
threshold $\eta$,
\begin{equation}
\frac{\sum_{i=1}^{r} \sigma_i^{2}}{\sum_{i=1}^{n} \sigma_i^{2}} \ge \eta,
\label{eq:energy}
\end{equation}
with $\eta = 1 - 10^{-7}$ used in cases B and C, $\eta = 1 - 10^{-14}$ in case A, and $\eta = 1 - 10^{-15}$ in case D. These values were determined heuristically.

\subsection{Time-Domain Prediction and Computational Cost}

After identification, the reduced transfer function can be
evaluated over the frequency grid corresponding to a prediction
interval. Let $U_{\mathrm{pred}}(j\omega_k)$ denote the FFT of
the input signal over that interval. The predicted output
spectrum is calculated as
\begin{equation}
    Y_{\mathrm{pred}}(j\omega_k)
    =
    H_r(j\omega_k)U_{\mathrm{pred}}(j\omega_k),
    \label{eq:predicted_output_spectrum}
\end{equation}
and the predicted time-domain response is obtained using the
inverse FFT. Since the time-domain signals are real, $H_r(j\omega_k)$ is evaluated only on the non-negative frequency bins and the predicted response is recovered with a real-valued inverse FFT, which enforces the conjugate symmetry. Because no zero-padding is applied, the inverse FFT implicitly treats the prediction window as periodic and any part of the response that has not decayed by the end of the window wraps around to its beginning, which appears as a small ripple before the excitation arrives in the time-domain results of Section~\ref{sec:validation}.

The prediction interval can be considerably longer than the
training window. A longer observation interval provides a finer
frequency grid, whereas the temporal sampling interval determines
the time resolution. The input spectrum and the reduced transfer
function must be evaluated on a consistent frequency grid.

The Loewner realization is not guaranteed to be stable: in all four
validation cases, between 2 and 9 of the $r$ generalized eigenvalues of
$(A_r,E_r)$ lie in the right half-plane. These poles do not compromise the
predictions reported here, because the reduced model is never integrated in
time; it is only evaluated on the imaginary axis through
$H_r(j\omega_k)$ in~\eqref{eq:predicted_output_spectrum}, where an unstable
pole contributes a bounded, if non-physical, frequency response. Time-stepping the descriptor system~\eqref{eq:descriptor_model} directly, or using the ROM outside the fitted band, would require these poles to be removed or reflected, which is left for future work together with passivity enforcement.

The computational benefit depends on the FOM size, the simulation duration, and the number of reduced-model evaluations. Identification adds the cost of assembling the Loewner matrices, $\mathcal{O}(n^2)$, and of their dense SVDs, $\mathcal{O}(n^3)$; for short simulations or small FOMs this may exceed the cost of simulating the FOM directly. The main benefit is therefore expected for large-scale problems, long prediction intervals, or repeated simulations with different excitations, where the identification cost is amortized.

Prediction accuracy is quantified by the normalized root-mean-square error
\begin{equation}
    \mathrm{NRMSE}
    =
    \frac{\| y_{\mathrm{pred}}-y_{\mathrm{ref}} \|_{2,\mathcal{I}}}
         {\| y_{\mathrm{ref}} \|_{2}},
    \label{eq:nrmse}
\end{equation}
where $\|\cdot\|_{2,\mathcal{I}}$ denotes the Euclidean norm over the samples
of an interval $\mathcal{I}$, evaluated separately over the training interval
$[0,t_{\mathrm{train}}]$ and the extrapolation interval
$(t_{\mathrm{train}},t_{\mathrm{full}}]$. The denominator is the norm of the
complete reference signal, so that the extrapolation error is not inflated
when the response has decayed.

\subsection{Numerical Implementation Details}

The frequency-selection criteria are important for obtaining a
reliable realization. Bins outside the excited frequency band are discarded because the input-output data may not contain sufficient information to identify the response at those frequencies. The band limits used for each case are listed in Table~\ref{tab:results}. The fit band may extend beyond the nominal excitation band where the spectral tails of the input still carry sufficient energy to resolve resonances of the structure, as in case~A, where resonances above 5~GHz are included.

The zero-frequency response requires particular care. When the
mean input is sufficiently far from zero, the ratio of the mean
output to the mean input provides an estimate of the DC gain:
\begin{equation}
    H(0)
    \approx
    \frac{\overline{y}_{\mathrm{train}}}
         {\overline{u}_{\mathrm{train}}}.
    \label{eq:dc_gain_estimate}
\end{equation}
This estimate may be unreliable if the training window is too
short, the signals have not reached the relevant steady-state
behavior, or the mean input is close to zero. In the present implementation, the zero-frequency bin of the predicted
spectrum in~\eqref{eq:predicted_output_spectrum} is replaced by this estimate.

Numerical care is also required when a right interpolation point
$\mu_i$ is very close to a left interpolation point $\lambda_j$.
In this case, the denominators in
\eqref{eq:loewner_matrix} and
\eqref{eq:shifted_loewner_matrix} become small, potentially making
the corresponding entries sensitive to round-off errors. However, the retained samples are assigned alternately to the right and left interpolation sets. Since both sets lie on the uniform FFT frequency grid,
$\mu_i \neq \lambda_j$ for all $i, j$, so the denominators in~\eqref{eq:loewner_matrix}
and~\eqref{eq:shifted_loewner_matrix} never vanish and no entries require special treatment.

\subsection{Identification and Prediction Algorithm}

The complete procedure is summarized in
Algorithm~\ref{alg:loewner}. The preceding subsections describe
the mathematical formulation and implementation choices, while
the algorithm lists the computational steps required to identify
the reduced model and evaluate its predictions.

\begin{algorithm}
\caption{Loewner-Based Identification and Time-Domain Prediction}
\label{alg:loewner}
\begin{algorithmic}[1]
\Require FOM input-output snapshots, sampling interval $\delta t$,
training duration $t_{\mathrm{train}}$, frequency-selection criteria,
and SVD truncation threshold.
\Ensure Reduced-order realization and predicted response
$y_{\mathrm{pred}}(t)$.

\State Extract the input and output signals over the training window.
\State Compute their FFTs.
\State Retain frequency bins satisfying the prescribed excitation criteria.
\State Estimate the transfer-function samples at the retained frequencies.
\State Partition the samples into left and right interpolation data.
\State Assemble the Loewner and shifted Loewner matrices, $\mathbb{L}$ and $\mathbb{L}_s$.
\State Compute the SVDs of $[\,\mathbb{L}\;\; \mathbb{L}_s\,]$ and
$[\,\mathbb{L};\, \mathbb{L}_s\,]$.
\State Select the reduced order $r$ using the prescribed truncation criterion.
\State Construct the reduced realization using the corresponding projection bases.
\State Evaluate $H_r(j\omega_k)$ on the prediction frequency grid.
\State Compute $Y_{\mathrm{pred}}(j\omega_k)
=H_r(j\omega_k)U_{\mathrm{pred}}(j\omega_k)$.
\State Apply the inverse FFT to obtain $y_{\mathrm{pred}}(t)$.
\end{algorithmic}
\end{algorithm}
\section{Validation}
\label{sec:validation}

\begin{figure}[t]
\centering
\includegraphics[width=\columnwidth]{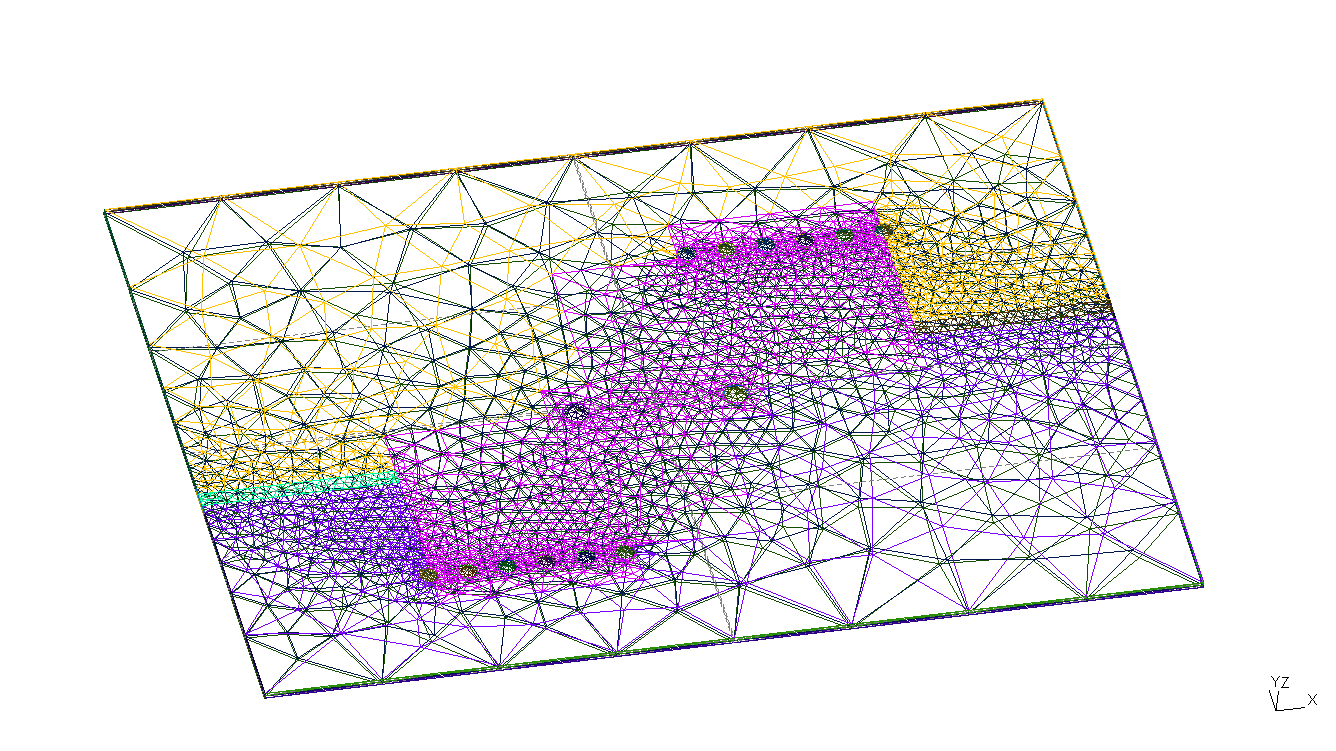}
\caption{Case A (QM-SIW filter): FOM discretization mesh of the quarter-mode SIW cavity structure used to generate the training and reference snapshots.}
\label{fig:tomassoni-model}
\end{figure}

\begin{figure}[t]
\centering
\includegraphics[width=\columnwidth]{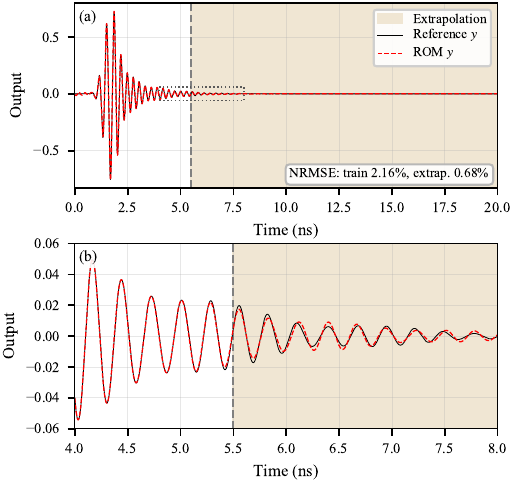}
\caption{Case A: time-domain reflection output, reference (FOM) versus Loewner ROM ($r=15$) trained on the first 5.5~ns; the extrapolation region is shaded. (a) First 20~ns of the 300~ns window. (b) Detail around the end of the training window (4--8~ns), marked by the dotted box in (a).}
\label{fig:caseA-time-refl}
\end{figure}

\begin{figure}[t]
\centering
\includegraphics[width=\columnwidth]{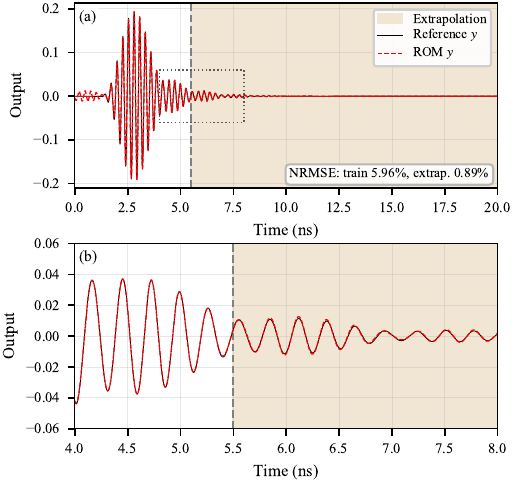}
\caption{Case A: time-domain transmission output, reference (FOM) versus Loewner ROM ($r=15$) trained on the first 5.5~ns; the extrapolation region is shaded. (a) First 20~ns of the 300~ns window. (b) Detail around the end of the training window (4--8~ns), marked by the dotted box in (a).}
\label{fig:caseA-time-trans}
\end{figure}

\begin{figure}[t]
\centering
\includegraphics[width=\columnwidth]{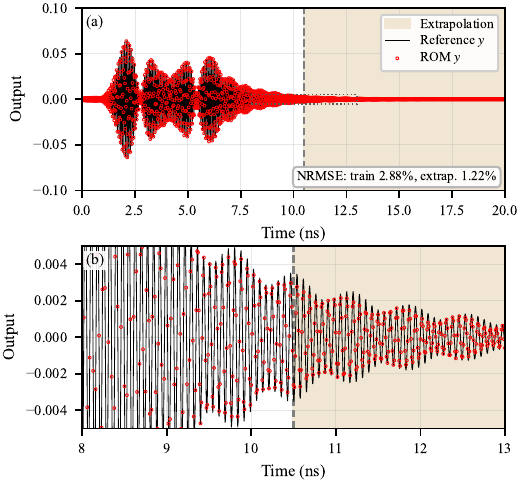}
\caption{Case B: time-domain reflection output, reference (FOM) versus Loewner ROM ($r=12$) trained on the first 10.5~ns; the extrapolation region is shaded and the ROM is shown as markers on a subset of the time samples for clarity. (a) First 20~ns of the 40~ns window. (b) Detail around the end of the training window (8--13~ns).}
\label{fig:caseB-time-refl}
\end{figure}

\begin{figure}[t]
\centering
\includegraphics[width=\columnwidth]{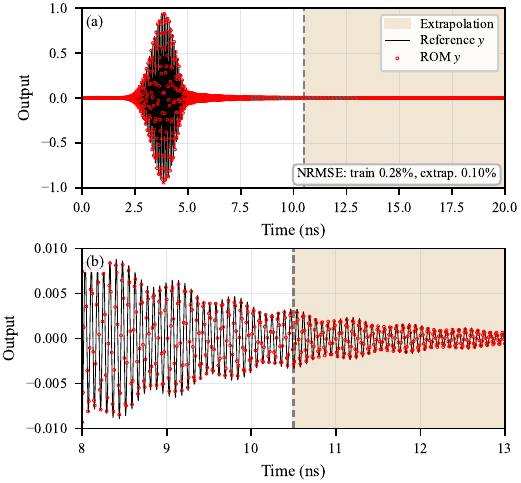}
\caption{Case B: time-domain transmission output, reference (FOM) versus Loewner ROM ($r=11$) trained on the first 10.5~ns; the extrapolation region is shaded and the ROM is shown as markers on a subset of the time samples for clarity. (a) First 20~ns of the 40~ns window. (b) Detail around the end of the training window (8--13~ns).}
\label{fig:caseB-time-trans}
\end{figure}

\begin{figure}[t]
\centering
\includegraphics[width=\columnwidth]{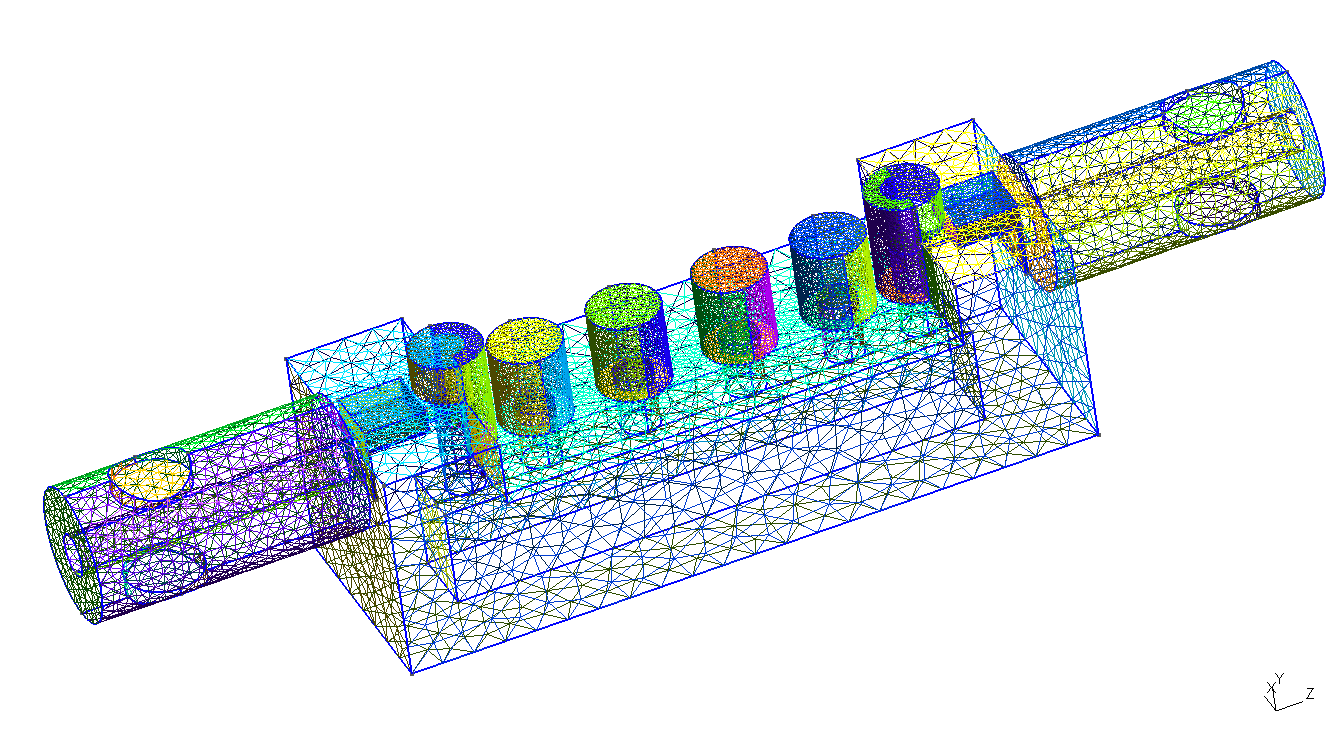}
\caption{Case C (broadband comb-line filter): FOM discretization mesh of the coaxial comb-line evanescent-mode structure used to generate the training and reference snapshots.}
\label{fig:bastioli-model}
\end{figure}

\begin{figure}[t]
\centering
\includegraphics[width=\columnwidth]{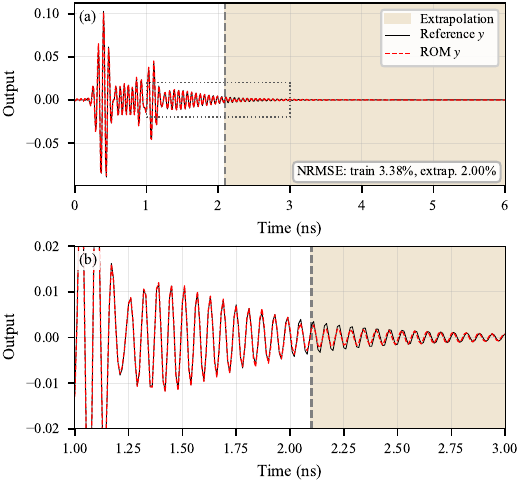}
\caption{Case C: time-domain reflection output, reference (FOM) versus Loewner ROM ($r=13$) trained on the first 2.1~ns; the extrapolation region is shaded. (a) First 6~ns of the 60~ns window. (b) Detail around the end of the training window (1--3~ns), marked by the dotted box in (a).}
\label{fig:caseC-time-refl}
\end{figure}

\begin{figure}[t]
\centering
\includegraphics[width=\columnwidth]{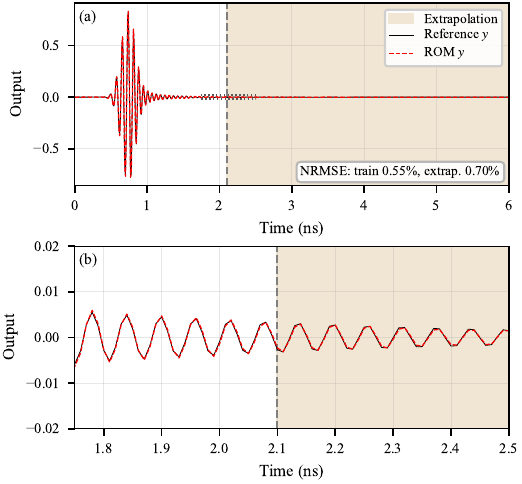}
\caption{Case C: time-domain transmission output, reference (FOM) versus Loewner ROM ($r=13$) trained on the first 2.1~ns; the extrapolation region is shaded. (a) First 6~ns of the 60~ns window. (b) Detail around the end of the training window (1.75--2.5~ns).}
\label{fig:caseC-time-trans}
\end{figure}

\begin{figure}[t]
\centering
\includegraphics[width=\columnwidth]{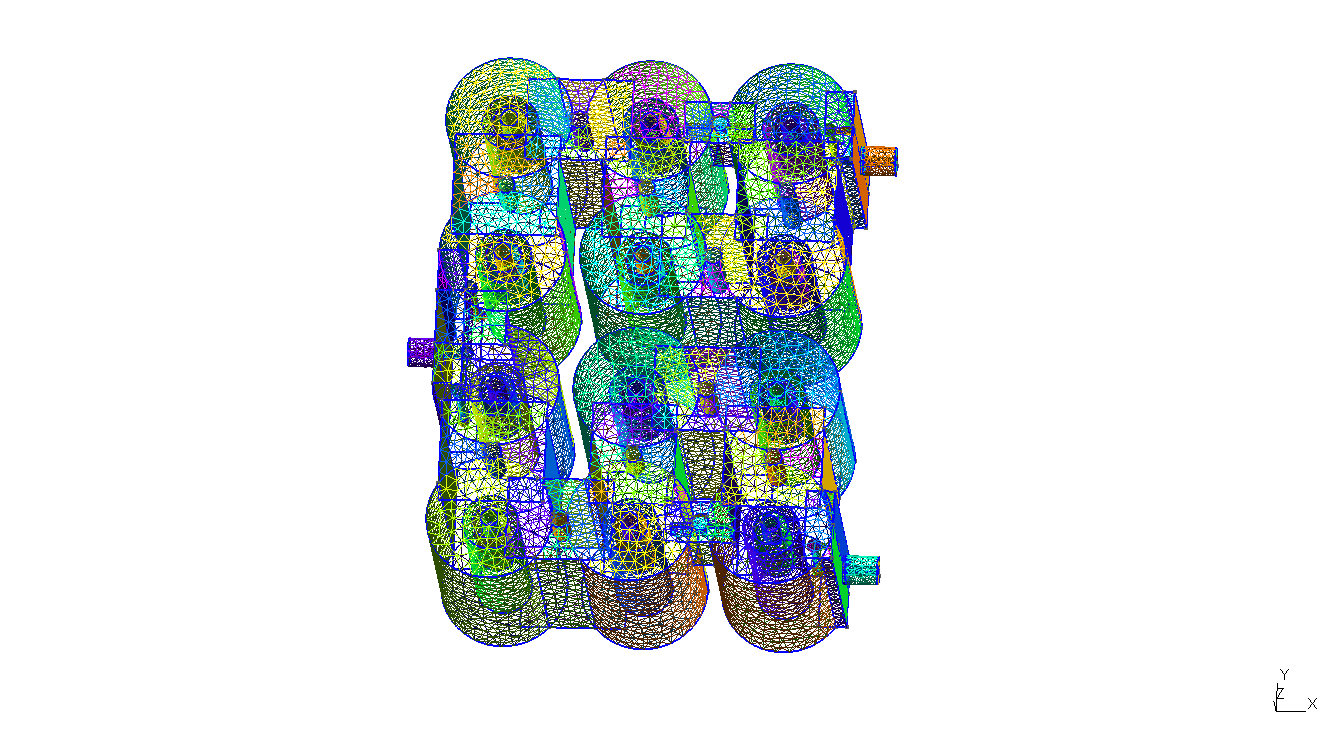}
\caption{Case D (diplexer): FOM discretization mesh of the comb-line-coupled diplexer structure, showing the common port and the two channel ports used to generate the training and reference snapshots.}
\label{fig:lamecki-model}
\end{figure}

\begin{figure}[t]
\centering
\includegraphics[width=\columnwidth]{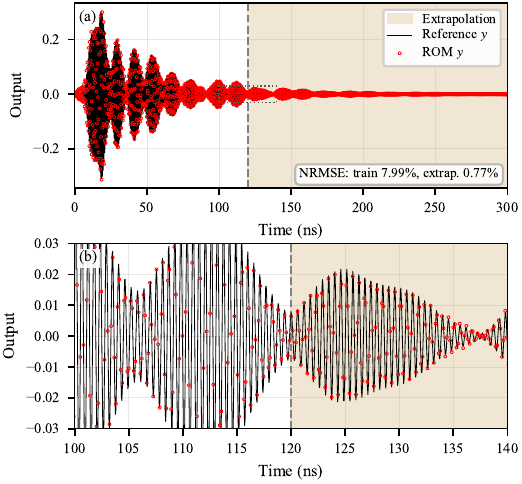}
\caption{Case D: time-domain reflection output at the common port, reference (FOM) versus Loewner ROM ($r=18$) trained on the first 120~ns; the extrapolation region is shaded and the ROM is shown as markers on a subset of the time samples for clarity. (a) Full 300~ns window. (b) Detail around the end of the training window (100--140~ns).}
\label{fig:caseD-time-refl}
\end{figure}

\begin{figure}[t]
\centering
\includegraphics[width=\columnwidth]{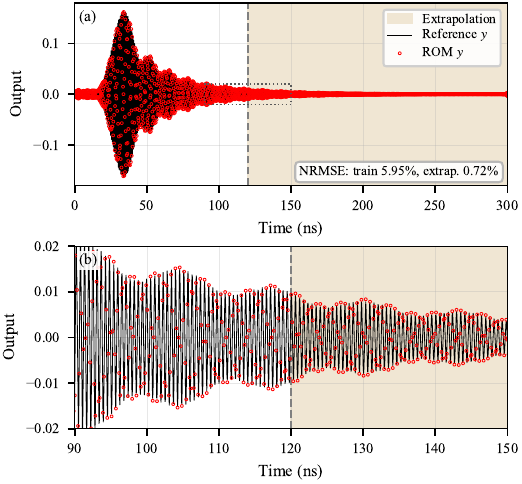}
\caption{Case D: time-domain transmission output to port~2, reference (FOM) versus Loewner ROM ($r=18$) trained on the first 120~ns; the extrapolation region is shaded and the ROM is shown as markers on a subset of the time samples for clarity. (a) Full 300~ns window. (b) Detail around the end of the training window (90--150~ns).}
\label{fig:caseD-time-t2}
\end{figure}

\begin{figure}[t]
\centering
\includegraphics[width=\columnwidth]{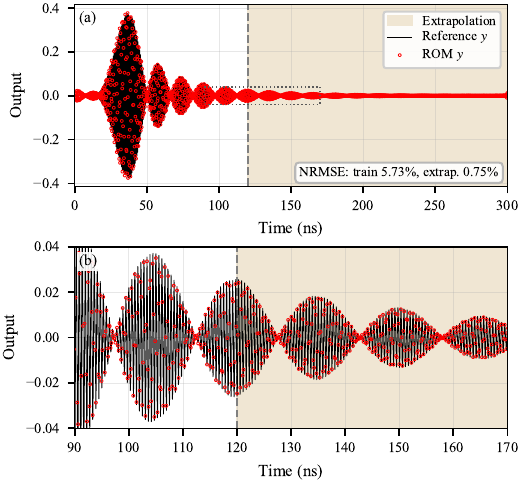}
\caption{Case D: time-domain transmission output to port~3, reference (FOM) versus Loewner ROM ($r=18$) trained on the first 120~ns; the extrapolation region is shaded and the ROM is shown as markers on a subset of the time samples for clarity. (a) Full 300~ns window. (b) Detail around the end of the training window (90--170~ns).}
\label{fig:caseD-time-t3}
\end{figure}

The method is validated on four microwave structures with FOM reference data generated with in-house FEM-TD/ROM-TD  solvers. Each structure is excited by a Gaussian pulse centered in its excitation band. The ROM is trained on a short initial window and then a response is extrapolated over the full simulated window; Table~\ref{tab:results} reports the resulting order reduction and cost, and Figs.~\ref{fig:tomassoni-model}--\ref{fig:caseD-time-t3} show, for each structure, the FOM discretization mesh used to generate the training snapshots, a comparison between the reference (FOM) and Loewner-predicted (ROM) scattering parameters $S_{ij}$, and the corresponding time-domain output of each port. Each $S_{ij}$ is identified as a separate single-input, single-output transfer function $H(s)$, with its own reduced order $r$, and its magnitude is plotted as $|H(j\omega)|$.

The same identification code (excited-band masking, DC correction, and frequency-domain simulation as described in Section~\ref{sec:method}) is applied unchanged to all four cases: no case requires a modified algorithm. The only inputs that differ from one case to the next are the raw FOM snapshots themselves and three user-set parameters: the fit band, the training-window length, and the SVD energy threshold $\eta$ (Table~\ref{tab:results} and Section~\ref{sec:method}). This uniformity is what makes the comparison across cases in Section~\ref{sec:sensitivity} meaningful: differences in ROM order and accuracy reflect the structures, their excitation, and these three parameters, rather than changes to the identification procedure itself. It also showcases the versatility of the pipeline, which can handle a variety of complex EM problems by adjusting only these three user-set parameters: the training window, the fit band and the energy threshold.

\subsection{Case A: QM-SIW Filter}

A quarter-mode substrate-integrated-waveguide (QM-SIW) cavity filter  \cite{moscato2016} excited over 1--5~GHz. From 275 snapshots ($\delta t=0.02$~ns) over a 5.5~ns training-window, the ROM reduces the FOM order from 136{,}352 to 15 and reproduces reflection and transmission over the full 300~ns window, cutting simulation cost from 48{,}200 to 95.6~ms per simulated nanosecond and the full-window execution time from about 4~h to under 5~min.

\begin{figure}[t]
\centering
\includegraphics[width=\columnwidth]{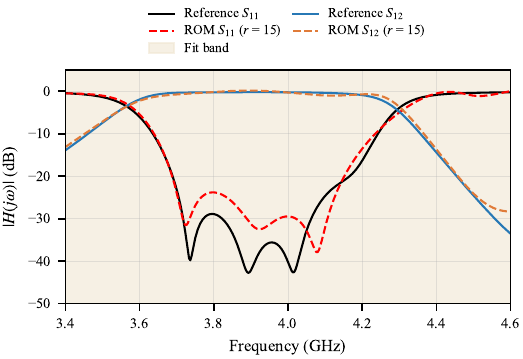}
\caption{Case A: reference (FOM) versus Loewner ROM ($r=15$) reflection $S_{11}$ and transmission $S_{12}$ for the reported 5.5~ns training window; the fit band is shaded.}
\label{fig:tomassoni-response}
\end{figure}

Fig.~\ref{fig:tomassoni-response} compares the reference (FOM) reflection and transmission, $S_{11}$ and $S_{12}$, against the Loewner ROM predictions over the 3.4--4.6~GHz passband region. The ROM trained on the 5.5~ns window (order $r=15$) closely follows the reference response across the band, reproducing both the passband flatness and the null structure of the quarter-mode SIW cavity filter from a training window less than 2\% the length of the full simulated window.

The three reflection minima visible between 3.7 and 4.1~GHz are the return-loss zeros of the coupled-cavity design, and their spacing (120--160~MHz on a 3.9~GHz center) is what the identification has to resolve and the deciding factor when determining the training-window. The ROM reproduces all three, the first two within about 30~MHz of the reference and the third shifted upward by about 60~MHz, but with a shallower in-band response: the predicted minima reach $-31.5$, $-32.5$ and $-38$~dB against $-40$, $-43$ and $-43$~dB for the reference, and the reflection level between them is about 5--6~dB higher than the reference. This is a benign error in engineering terms, since the predicted return loss remains better than 23~dB between 3.7 and 4.1~GHz, and it is characteristic of the method: a pointwise interpolant is constrained only at the retained frequency samples, and in a region where $|H|$ varies by tens of dB between adjacent samples the depth of the minimum between them is essentially unconstrained.

The transmission response is reproduced almost exactly across the passband, with the ROM and FOM curves within 1~dB of each other between 3.5 and 4.25~GHz. The two begin to diverge only above about 4.5~GHz, on the upper skirt: at 4.6~GHz the predicted $S_{12}$ is $-28$~dB against $-33.5$~dB for the reference. This case shows a moderate-$Q$, well-separated pole structure over a single 1--5~GHz band. It is also the one with the largest extrapolation ratio, since the 5.5~ns training-window is used to predict a 300~ns response, a factor of about 54.

\subsection{Case B: 3-D FSS}

A three-dimensional, fully metallic, dual-polarization frequency-selective surface (FSS) \cite{parellada2024} excited over 11--15~GHz. Training on 210 snapshots ($\delta t=0.05$~ns, 10.5~ns window) reduces the FOM order from 285{,}974 to 12 for the reflection and 11 for the transmission, and the per-nanosecond cost from 51{,}800 to 8.35~ms, extrapolating accurately to the 40~ns full window.

\begin{figure}[t]
\centering
\includegraphics[width=\columnwidth]{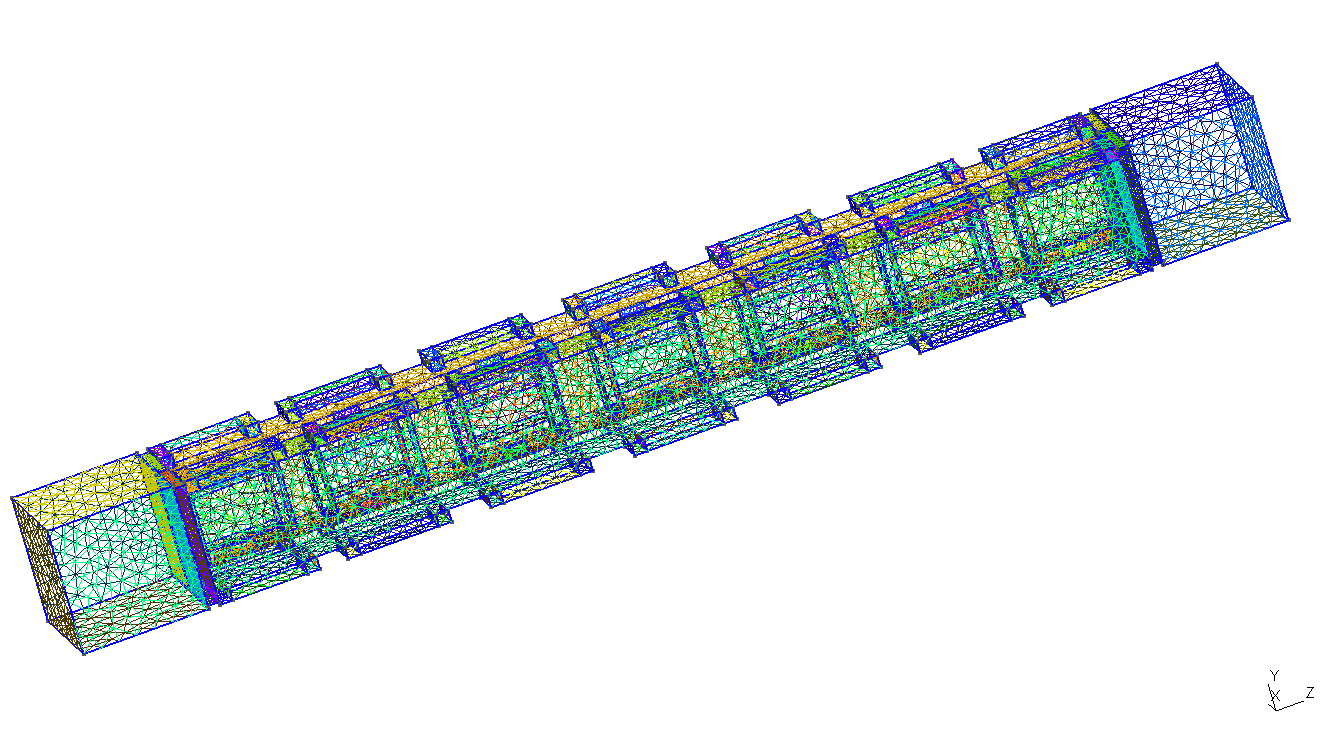}
\caption{Case B (3-D FSS): FOM discretization mesh of the fully metallic, dual-polarization frequency-selective-surface unit cell used to generate the training and reference snapshots.}
\label{fig:molero-model}
\end{figure}

\begin{figure}[t]
\centering
\includegraphics[width=\columnwidth]{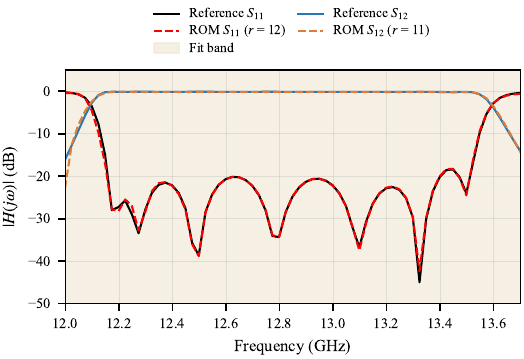}
\caption{Case B: reference (FOM) versus Loewner ROM reflection $S_{11}$ ($r=12$) and transmission $S_{12}$ ($r=11$) for the reported 10.5~ns training window; the fit band is shaded.}
\label{fig:molero-response}
\end{figure}

Fig.~\ref{fig:molero-response} compares the reference (FOM) reflection and transmission, $S_{11}$ and $S_{12}$, against the Loewner ROM predictions over the 12--13.7~GHz frequency window shown. The ROMs trained on the 10.5~ns window ($r=12$ for $S_{11}$ and $r=11$ for $S_{12}$) are in great accord with the reference curves across the whole band despite retaining only 11--12 states out of a FOM of 285{,}974 degrees of freedom.

The reflection curves are a close match over the entire 12.2--13.5~GHz interval, and the ROM captures the seven expected zeroes of the frequency selective surface that correspond to the H-shaped resonators shown in Fig.~\ref{fig:molero-model}. Each of these minima is a separate return-loss zero produced by the coupled-resonator behavior of the dual-polarization unit cell, and reproducing them individually rather than as an averaged envelope is proof that the identification has accurately resolved the underlying poles instead of merely smoothing over them.

The main area of disagreement between predicted and reference data is confined to the lower edge of the fitted band. Between 12.0 and 12.3~GHz the predicted transmission departs from the reference by up to about 1.3~dB. This is the band edge, where the excitation spectrum has begun to fall away. Notably, case B combines the smallest extrapolation ratio of the four cases (40~ns from a 10.5~ns window, a factor of about 4) with the highest in-band fidelity. This supports the reasoning that accuracy in replicating the underlying physics of the system is governed by how well the training-window resolves the pole structure and not by how far the model is subsequently extrapolated. Taking this into consideration, accuracy in the lower-frequency regime of the passband could be improved by increasing the training-window.

\subsection{Case C: Broadband Combline Filter}

A coaxial comb-line evanescent-mode filter with a resonant input \cite{bastioli2024}, excited broadband over 1--25~GHz. A 2.1~ns training-window (210 snapshots, $\delta t=0.01$~ns) yields a 13th-order ROM from a 371{,}220th-order FOM, lowering the cost per nanosecond from 366{,}600 to 4.75~ms over the 60~ns full window.

\begin{figure}[t]
\centering
\includegraphics[width=\columnwidth]{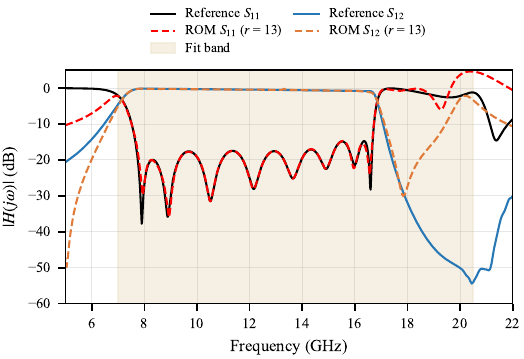}
\caption{Case C: reference (FOM) versus Loewner ROM ($r=13$) reflection $S_{11}$ and transmission $S_{12}$ for the reported 2.1~ns training window; the fit band is shaded.}
\label{fig:bastioli-response}
\end{figure}

As shown in the previous cases, Fig.~\ref{fig:bastioli-response} compares the reference (FOM) reflection and transmission, $S_{11}$ and $S_{12}$, against the Loewner ROM predictions over the 7--20.5~GHz fit band. The 13th-order ROM (2.1~ns training-window) reproduces the eight-ripple passband and the transition into both stopbands almost exactly between roughly 7 and 17~GHz (the widest relative bandwidth of the four validated cases).

The eight reflection minima, spread from roughly 7.9 to 16.6~GHz, are tracked individually, and over the central 8--15~GHz region the predicted and reference curves agree to within about 3~dB. Two deviations can be observed, both at the edges of the fit band. First, below its 7~GHz lower edge, where no transfer-function samples constrain the model, the predicted transmission rolls off much faster than the reference ($-50$~dB against $-20$~dB at 5~GHz), and the predicted reflection drops to about $-10$~dB where the reference remains close to 0~dB. Second, above 18~GHz the predicted transmission departs from the reference: while the reference rolls off from $-32$~dB at 18~GHz to $-54$~dB at 20.5~GHz, the prediction rises back to $-15$~dB at 19~GHz and about $-3$~dB at 20~GHz, accompanied by a predicted reflection that exceeds 0~dB by up to 4.6~dB. Although this second region lies inside the fit band, the transmission there is 30--55~dB below the passband, beyond the dynamic range that a model of this order reproduces, as also observed for cases~A and~D; the ROM places numerical poles there that have no physical counterpart and that yield a non-passive response. The ROM is therefore reliable between roughly 7 and 17~GHz, and should not be used outside this range.

\subsection{Case D: Diplexer}

A higher-order 3-D finite-element microwave diplexer \cite{lamecki2014}, excited over 1.4--2.2~GHz. With the longest reported training-window (120~ns, 2400 snapshots, $\delta t=0.05$~ns), the ROM reduces the FOM order from 478{,}778 to 18 and the per-nanosecond cost from 102{,}900 to 1.04~ms, correctly separating the Tx and Rx channel responses over the full 300~ns window.

\begin{figure}[t]
\centering
\includegraphics[width=\columnwidth]{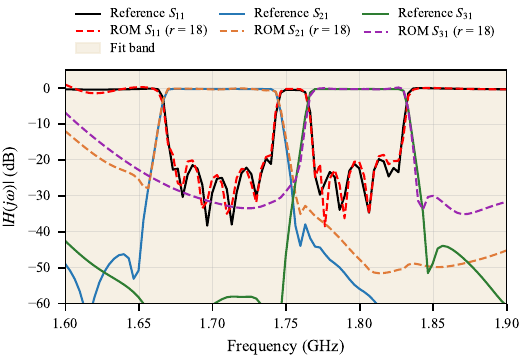}
\caption{Case D: reference (FOM) versus Loewner ROM ($r=18$) reflection $S_{11}$ and channel transmissions $S_{21}$, $S_{31}$ for the reported 120~ns training window; the fit band is shaded.}
\label{fig:lamecki-response}
\end{figure}

Fig.~\ref{fig:lamecki-response} compares the reference (FOM) reflection and transmission, $S_{11}$, $S_{21}$, and $S_{31}$, against the Loewner ROM predictions over the 1.6--1.9~GHz band spanning both channels. The 18th-order ROM (120~ns training-window) matches the reference passband edges and the shared-junction transition between the two closely spaced Tx/Rx channels, correctly separating the two transmission paths despite the tight, tens-of-MHz channel spacing that makes this diplexer the most demanding of the four cases: its Tx and Rx passbands share a common junction, and a training-window that stops before both channels have rung down measurably would leave the Loewner interpolation under-determined near the shared junction resonance. 

The general behavior of the diplexer is adequately captured. Both channel passbands appear at the correct frequencies, the crossover near 1.75~GHz is placed correctly, and the reflection ripple pattern within each passband is reproduced in shape, although the shallow minimum near 1.82~GHz at the upper edge of the second channel is not resolved. However, within the passbands individual reflection minima are matched only to within several dB and occasionally much more: near 1.70~GHz the predicted minima are 3--5~dB shallower than the reference ($-33.5$ against $-38.3$~dB at 1.696~GHz), whereas near 1.775~GHz the predicted zero is about 9~dB deeper ($-38.5$ against $-29.7$~dB). This is the same unconstrained-minimum behavior previously noted for case A, but more pronounced here, since the ripples are narrower relative to the frequency-sample spacing. The deviations in the reflection are related to the complexity of this particular case. The narrow space between the zeroes of the transfer function enforces a bigger training-window when compared to previous cases, this is caused by the resolution with which the ROM captures the physics of the system being limited by the separation of the frequency bins of the FFT performed on the datasets, which is determined by the training-window. This idea is further developed in Section~\ref{sec:sensitivity}.

Once again, a limitation is seen in the out-of-band isolation. Within the opposite channel's passband the reference transmissions fall to between $-44$ and $-69$~dB, whereas the predicted $S_{21}$ levels off between $-37$ and $-52$~dB and the predicted $S_{31}$ between $-28$ and $-34$~dB, and neither follows the reference further down. The practical implication is specific and worth stating plainly: the current implementation of a Loewner ROM of this order reproduces passband shape, insertion loss and band edges faithfully; but should not be relied upon to predict deep stopband rejection or channel isolation figures. Comparable departures in the low-level regions of the response are visible in case C's transmission of Fig.~\ref{fig:bastioli-response} near the edges of the fit band and, more mildly, in case A's upper skirt, so this is a property of the system identification rather than of the diplexer.

\begin{table}[t]
\centering
\caption{Summary of Validation Results}
\label{tab:results}
\scriptsize
\begin{tabular}{@{}lcccc@{}}
\toprule
\textbf{Metric} & \textbf{Case A} & \textbf{Case B} & \textbf{Case C} & \textbf{Case D} \\
\midrule
Excitation band (GHz) & 1--5 & 11--15 & 1--25 & 1.4--2.2 \\
Fit band (GHz) & 0--6 & 11.7--14 & 7--20.5 & 1.6--1.9 \\
Snapshots / $\delta t$ (ns) & 275/0.02 & 210/0.05 & 210/0.01 & 2400/0.05 \\
FOM order ($N$) & 136,352 & 285,974 & 371,220 & 478,778 \\
ROM order ($r$) & 15 & 12 / 11 & 13 & 18 \\
Training window (ns) & 0 - 5.5 & 0 - 10.5 & 0 - 2.1 & 0 - 120 \\
Full window (ns) & 300 & 40 & 60 & 300 \\
FOM cost (ms/ns) & 48,200 & 51,800 & 366,600 & 102,900 \\
ROM cost (ms/ns) & 95.6 & 8.35 & 4.75 & 1.04 \\
Per-ns eval. speed-up & 504$\times$ & 6,204$\times$ & 77,179$\times$ & 98,942$\times$ \\
Full-run, FOM$\to$ROM & 4h$\to$4m53s & 34m$\to$9m & 6h$\to$13m & 8h34m$\to$3h25m \\
Max.\ NRMSE, train (\%) & 5.96 & 2.88 & 3.38 & 7.99 \\
Max.\ NRMSE, extrap.\ (\%) & 0.89 & 1.22 & 2.00 & 0.77 \\
\bottomrule
\end{tabular}
\end{table}

Across the four cases, ROM order grows only mildly with FOM complexity (from 11 for case B to 18 for case D) while the per-ns evaluation speed-up spans more than two orders of magnitude, from 504$\times$ for case A to nearly $10^{5}\times$ for cases C and D. These gains confirm that the Loewner ROM is quite powerful and useful when the FOM simulation is both large and required over long physical times. The training-window sweep summarized in Table~\ref{tab:sensitivity} further shows a consistent pattern: accuracy decreases as the training-window shortens and the model becomes unable to resolve the physics of the system, making under-training easy to diagnose directly from the predicted transfer function, which could potentially lead to a non-heuristic way to determine the training-window.

Read together, the four cases separate the factors that govern accuracy from those that govern cost. The extrapolation ratio seems to drive the saving but not the error: cases A and B sit at opposite ends of that ratio (54$\times$ against 4$\times$) yet case B is the more accurate fit of the two. What does correlate with error is the amount of usable information per frequency sample, or in other words, how strongly each pole is excited within the training-window. Narrowband strongly excited structures such as the one in case B are reproduced almost exactly with very low order; broadband excitation (case C) thins the energy available per bin, and closely spaced channels near a shared junction (case D) demand resolution the retained samples cannot supply, which relates to the length of the training-window discussed in Section~\ref{sec:sensitivity}. All four cases share a common ceiling on achievable dynamic range: agreement is excellent down to roughly 25--40~dB below the passband and degrades below that, so these ROMs are well suited to predicting passband behavior, band edges and insertion loss, and poorly suited to predicting deep rejection or isolation as well as out-of-band behavior.

\subsection{Sensitivity to Training-Window Length}
\label{sec:sensitivity}

Table~\ref{tab:sensitivity} collects, for each case, three training-window lengths swept during development and the resulting ROM order $r$, giving a direct, empirical view of how the identified order depends on how much of the transient is observed. The shortest window in each sweep consistently produces the lowest order and the least accurate fit, typically failing first as a spurious resonance or an unbounded excursion near a band edge rather than as a uniformly degraded curve.

\begin{figure}[t]
\centering
\includegraphics[width=\columnwidth]{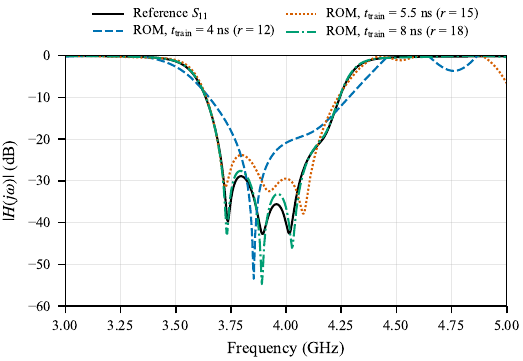}
\caption{Case A, reflection $S_{11}$: reference response and Loewner ROM predictions for training windows of 4~ns ($r=12$), 5.5~ns ($r=15$, reported case) and 8~ns ($r=18$).}
\label{fig:tomassoni-sweep}
\end{figure}

Fig.~\ref{fig:tomassoni-sweep} shows this dependence between training-window length and accuracy of the ROM for case A, superimposing the reference reflection on the ROM prediction obtained from training-windows of 4~ns ($r=12$, blue dashed), 5.5~ns ($r=15$, orange dots) and 8~ns ($r=18$, green dash-dotted). The comparison separates two distinct effects of the window length. Inside the excitation band, all three models already replicate the behavior of the filter with different degrees of accuracy and place the reflection nulls near 3.8--4.0~GHz at essentially the correct frequencies: the pole locations (the physics of the resonant cavity) are recovered even from the shortest window, because the Loewner interpolation approximates sampled transfer-function values pointwise instead of fitting a rational function of arbitrarily chosen order. What the longer windows improve is the \emph{depth and separation} of those nulls: the 4 ns model (r = 12) merges the three zeros into a single deep notch near 3.88 GHz, the 5.5 ns model resolves them individually but 8–10 dB shallower than the reference, and only the 8 ns model tracks the reference minima to within a few dB. The closely spaced poles are only distinguishable once the transient has been observed long enough for their beat to develop and the separation between the frequency bins can properly resolve them, so the window length effectively sets the frequency resolution with which nearby resonances can be separated.

The second effect is visible at the edges of the fit band, below roughly
1~GHz and above 5~GHz, where the input spectrum decays and the
transfer-function samples become increasingly unreliable. There, each
identified order places its own spurious poles, with the artifact frequency
shifting from case to case rather than converging. The practical reading of
Fig.~\ref{fig:tomassoni-sweep} is therefore that a training-window must be
long enough to resolve the closely spaced poles that define the response, but that the resulting ROM should only be trusted over the band where the
excitation actually deposited energy. The fact that under-training manifests
as a localized, obviously non-physical feature rather than a uniformly
degraded curve is what makes the sweep usable as a diagnostic.

\begin{table}[t]
\centering
\caption{Training-Window Sensitivity (Window $\to$ ROM Order)}
\label{tab:sensitivity}
\scriptsize
\begin{tabular}{@{}lccc@{}}
\toprule
\textbf{Case} & \textbf{Short} & \textbf{Reported} & \textbf{Long} \\
\midrule
Case A & 4 ns $\to$ $r{=}12$ & 5.5 ns $\to$ $r{=}15$ & 8 ns $\to$ $r{=}18$ \\
Case B & 4 ns $\to$ $r{=}5$ & 10.5 ns $\to$ $r{=}12/11$ & 40 ns $\to$ $r{=}22$ \\
Case C & 1.5 ns $\to$ $r{=}11$ & 2.1 ns $\to$ $r{=}13$ & 10 ns $\to$ $r{=}44$ \\
Case D & 90 ns $\to$ $r{=}13$ & 120 ns $\to$ $r{=}18$ & 150 ns $\to$ $r{=}20$ \\
\bottomrule
\end{tabular}
\end{table}

A practical consequence for deploying Algorithm~1 on a new structure is that the training-window sweep itself is inexpensive, since steps 2--7 cost little next to the FOM run, and can therefore be used as a diagnostic: repeating the identification for two or three candidate window lengths and checking that the predicted transfer function stabilizes is a cheap, purely data-driven substitute for an a priori estimate of how long the training-window needs to be.

\section{Conclusions}

The time-domain Loewner framework, formalized here as Algorithm~1 and validated on four practical EM structures, was successfully applied to reduced-order modeling directly from time-domain simulation snapshots. Across the four validation cases, the resulting ROMs are three to four orders of magnitude lower in order than the corresponding FOMs and reproduce the reflection and transmission response with high fidelity while extrapolating well beyond the training window (up to 54 times longer for case A and 29 times longer for case C) cutting per-nanosecond evaluation cost by up to five orders of magnitude and end-to-end run time from hours to minutes in two out of the four cases of study. The fidelity is highest in the passband and degrades below roughly 25--40~dB of dynamic range, so the reduced models are best used to predict passband response, band edges and insertion loss rather than deep stopband rejection.

Future work will focus on improving the numerical stability and transient behavior of the reduced models (in particular the sensitivity to training-window length observed in Section~\ref{sec:validation} as well as the enforcing of passivity to ensure good out-of-band behavior) extending the framework to predict the output for new, previously unseen inputs once the transfer function of a system has been identified, and generalizing the approach to non-linear systems through Koopman-operator-based methods such as dynamic mode decomposition (DMD). Larger-scale and more demanding EM problems, such as full-aircraft lightning-strike, are also a promising next target for this data-driven reduction strategy.

More broadly, the fact that Algorithm~1 requires nothing beyond time-domain excitation and response snapshots means it can be layered on top of essentially any existing time-domain EM workflow without modifying the underlying solver, making it a practical, low-effort first step toward reduced-order modeling for groups that do not have access to, or do not wish to modify, the internal system matrices of their full-order simulations.

A limitation of the present validation is that all four cases use synthetic FOM data generated by full-wave time-domain solvers rather than measurements, so the Loewner ROM is being asked to reproduce simulated, essentially noise-free transients. Measured data would add sensor noise and calibration error on top of the effects already visible in the figures above, and the masking safeguards in Algorithm~1 would need to be tuned accordingly. Likewise, the training-window sweeps in Section~\ref{sec:sensitivity} were run case by case rather than through a single automated rule for choosing $t_{\mathrm{train}}$, so translating the qualitative diagnostic proposed above into a quantitative stopping criterion is left for future work, as is the implementation of a method to ensure passivity of the ROM which would further improve robustness of the method.

Within these limits, the consistency of the results across four structures with very different topologies, bandwidths, and FOM sizes: a cavity filter, a frequency-selective surface, a comb-line filter, and a diplexer; suggests that the training-window sensitivity documented in Section~\ref{sec:sensitivity} is a property of the Loewner identification itself rather than of any solver or geometry, which is the main reason Algorithm~1 is reported here in an implementation-level form rather than only as an abstract data-driven realization procedure. Taken together, Algorithm~1, the explicit Loewner and shifted-Loewner matrices of Section~\ref{sec:method}, and the four validated cases of Section~\ref{sec:validation} are intended to make the method directly reproducible: a reader with access to time-domain excitation and response snapshots for their own structure can follow the steps of Algorithm~1 as written, without any need to consult the underlying interpolation-theoretic literature first.
\section*{Acknowledgment}
The authors used Claude (Anthropic) during the preparation of this manuscript to polish grammar and vocabulary, and to identify and correct inconsistencies between the text, tables and figures introduced during revision. All content was reviewed and verified by the authors, who take full responsibility for the final manuscript.

\end{document}